\documentstyle[12pt]{article}
\begin{document}

\vspace{2cm}

\begin{center}
    {\Large  {\bf EFFECT OF THE RELATIVISTIC SPIN ROTATION
      FOR ONE- AND TWO-PARTICLE SPIN STATES }}
\vskip 0.15in
        {\bf R. Lednicky, V.L. Lyuboshitz, V.V. Lyuboshitz }\\
{\it Joint Institute for Nuclear Research}\\
\end{center}
\begin{abstract}
The effect of the relativistic spin rotation, conditioned by the
setting of the spin in the rest frame of a particle and by the
noncommutativity of the Lorentz transformations along noncolinear
directions, is discussed. In connection with this, the Thomas
precession of the spin polarization vector at the motion of a
particle along a curvilinear trajectory is considered.  The
transformations of the correlation tensor components for a system
of two spin-1/2 particles at the transition from the c.m.s. of the
particle pair to the laboratory frame are investigated.
When the particle laboratory velocities are not
colinear, the relativistic spin rotation angles
for these particles are different.
As a result, the relative fractions
of the singlet and triplet states in the relativistic system of two
free spin-1/2 particles with a nonzero vector of relative
momentum  depend on the concrete frame in which this
two-particle system is analyzed.
\end{abstract}

\section{Introduction}
\noindent

In modern relativistic theory of reactions, the S-matrix is
parameterized on the basis of the formalism of inhomogeneous
Lorentz group [1-7].
This is similar
to the nonrelativistic theory with the essential modification -
the relativistic spin rotation.  Since, in this formalism,
the spin
state of a particle is  set in its rest frame,
the {\it concrete} description of this state
depends
on the frame from which the Lorentz transformation to the rest
frame is performed.
This is the essence of the effect of relativistic spin
rotation  at the transition from one frame to another [5-7].
The  relativistic spin rotation is a purely kinematical effect
conditioned by the setting of the spin of a particle in its rest
frame and by the
additional rotation of the spatial axes at the
successive Lorentz transformations along noncolinear directions,
the latter leading to the nontransitivity
of the parallelism in the theory of relativity (see \cite{shi99}
and references therein).

For two free particles, the total spin results from the addition
of the particle spins, defined in the respective rest frames,
according  to the usual angular momentum addition rules.
The operator of the total angular momentum is represented
by a sum of the operators of total spin and orbital angular momentum,
both commuting with the free Hamiltonian. The price for this quasi
nonrelativistic description is the lost relation of the orbital
angular momentum operator with the usual field coordinate.
Instead, this operator is related with the
c.m.s. coordinate introduced by Pryce \cite{pry49},
Newton and Wigner \cite{nw49}
and, within the Dirac
theory - by Foldy and Wouthuysen \cite{fw50}
(see also papers \cite{fol56,cs58,rit61}).

In the case of a nonzero vector of relative velocity, the angles
of the relativistic rotation of the spins of the two particles
are different - they depend on the frame  from which
the Lorentz transformations to the particle rest frames are
performed. This leads to the frame dependence of the
fractions of the states with different values of the total
spin (in particular, the singlet and triplet fractions
for a system of two spin-1/2 particles).
Thus, due to the frame dependence of the difference of the
spin rotation angles,
the square of the total spin
of a system of two free particles
with a nonzero vector of relative velocity
is \underline{not a Lorentz invariant}.

In Sections 2 and 3 a brief review of the
consequences of the relativistic spin rotation for one-particle
spin states is given; the connection of the effect of the
relativistic spin rotation with the kinematical spatial rotations
at successive Lorentz transformations is analyzed.
In Section 4, the relation between
the relativistic spin rotation  and the Thomas precession of the
internal angular momentum at the motion  of a particle along a
curvilinear trajectory is discussed.
The precession of the spin
polarization vector of a relativistic particle in the magnetic and
electric fields is considered in Section 5.
In Sections 6 and 7
the effect of the relativistic spin rotation for the states of
two free particles with a nonzero vector of relative momentum is
considered in detail.
The results are summarized in Section 8.

\section{Relativistic spatial rotation}
\noindent

Let $ A $ be any four-vector set in a frame $K$:
$ A = \{ A_0, {\bf A}\},~A^2 = A_0^2 - {\bf A}^2$.
The components of this four-vector in another frame $K'$,
having parallel respective spatial axes and moving with velocity
${\bf v}$ and Lorentz factor $\gamma=(1-{\bf v}^2/c^2)^{-1/2}$
($c$ is the velocity of light)
with respect to the frame $K$, are given by the
Lorentz transformation:
\begin{equation}
\label{1}
{\bf A}' = {\bf A}-\frac{{\bf v}}{c}\frac{\gamma }{\gamma +1}
(A_0+A_0'),~~A_0'=\gamma\left(A_0-\frac{{\bf v}}{c}{\bf A}
\right),
\end{equation}
$A'^2 = A^2$.
We will denote this transformation as $A'=L({\bf v})A$.\footnote{
In accordance with the principle of relativity, the parallelism
of the coordinate axes requires that the velocity vector
of the frame $K$ with respect to the frame $K'$ is $-{\bf v}$.
One can check that the inverse transformation
$L^{-1}({\bf v})=L(-{\bf v})$, i.e. $A=L(-{\bf v})A'$.
The above definition of parallelism is not unique in the case
when the direction of the frame velocity coincides with one
of the coordinate axes. In such a case, the observers in the
frames $K$ and $K'$ have to specify the plane containing their
relative velocity vector and one of the remaining two axes.
This can be easily
done, e.g., with the help of a light signal.
}
Let us further consider another frame $K_0$
associated with a physical object
moving with the velocity
${\bf v_0}$
and Lorentz factor $\gamma_0$, having
parallel spatial axes with respect to the
frame $K$.
The same object moves with the velocity
${\bf v_0}'$
and Lorentz factor $\gamma_0'$ with respect to the frame $K'$ -
we denote $\widetilde{K}_0$ the associated frame having
parallel spatial axes with respect to the frame $K'$.
We will see that, generally, $\widetilde{K}_0\ne K_0$:
the parallel axes of the
frames $K'$ and $K$, $K_0$ and $K$ do not imply the parallel axes
of the frames $K_0$ and $K'$. The axes of all the three frames
could be mutually parallel if only their velocities were colinear.

Considering a particle of four-velocity $u=p/m$
(p is the particle four-momentum and m is its mass)
at rest in the frame $\widetilde{K}_0$ or $K_0$:
$u^*=\{c,0,0,0\}$,
we can transform it to the frame $K'$:
$u'=L(-{\bf v}_0')u^*=\{\gamma_0'c,\gamma_0'{\bf v}_0'\}$,
and then to the frame $K$:
$u=L(-{\bf v})L(-{\bf v}_0')u^*$.
Comparing the result of the successive transformations
with that of the direct one:
$u=L(-{\bf v}_0)u^*=\{\gamma_0c,\gamma_0{\bf v}_0\}$,
we get the vector form of the law of the relativistic addition
of the velocities ${\bf v}_0'$ and ${\bf v}$:
\begin{equation}
{\bf v}_0 =\frac{\gamma \gamma_0'}{\gamma_0} \left [{\bf
v}_0'\frac{1}{\gamma} + {\bf v} \frac{\gamma_0' + \gamma_0}{\gamma_0'
(\gamma + 1)}\right ]=\left[
\frac{{\bf
  v}_0'}{\gamma} + {\bf v} \left (1 + \frac{{\bf v \,v}_0'} {{\bf
v}^2}\frac {\gamma - 1}{\gamma} \right )\right]
\left( 1 + \frac{{\bf
v\,v}_0'}{c^2}\right)^{-1}
.
\end{equation}
The second equality follows from the transformation
of the Lorentz factors (see the second equation in Eq.~(\ref{1})):
$$
\gamma_0 = \gamma \gamma_0' \left ( 1 +
\frac{{\bf v\,v}_0'}{c^2}\right).
$$

At the relativistic addition of velocities in the reverse order:
$u=L(-{\bf v}_0')L(-{\bf v})u^*$,
we obtain
 \begin{equation}
 \widetilde {{\bf v}}_0 =
 \frac{\gamma \gamma_0'}{\gamma_0} \left [{\bf v}\frac{1}
{\gamma_0'} +
{\bf v}_0' \frac{\gamma + \gamma_0}{\gamma (\gamma_0' + 1)}\right ]
= \left[
 \frac{{\bf
  v}}{\gamma_0'} + {\bf v}_0' \left (1 + \frac{{\bf v \,v}_0'} {{\bf
v}_0'^2}\frac {\gamma_0' - 1}{\gamma_0'} \right )\right]
\left( 1 + \frac{{\bf
v\,v}_0'}{c^2}\right)^{-1}
.
\end{equation}
For colinear
vectors ${\bf v}$ and ${\bf v}_0'$, the results of
the addition of velocities in the direct and reverse order coincide.
In the case of noncolinear  velocities ${\bf v}$ and ${\bf v}_0'$,
we have ${\bf v}_0 \ne \widetilde {\bf v}_0$
(but always $|{\bf v}_0| =
|\widetilde {{\bf v}}_0|, \gamma_0 = \widetilde{\gamma}_0$).
Thus, the symmetry in the addition of velocities, in general,
is absent.
The successive Lorentz transformations along noncolinear directions
are noncommutative; the pure Lorentz transformations do not form a
group.

Let us now compare the results of direct ($K\rightarrow K_0$)
and successive ($K\rightarrow K'\rightarrow {\widetilde K}_0$)
Lorentz transformations of an arbitrary four-vector $ A $.
The direct Lorentz transformation of this four-vector from the
frame $K$ to the frame $K_0$ gives
 \begin{equation}
   A^* = L( {\bf v}_0 )\, A,
\end{equation}
where ${\bf v}_0$ is the velocity of the frame $K_0$ in the
frame $K$,
determined by the relativistic addition of the velocities
${\bf v}_0'$ and ${\bf v}$.
The successive Lorentz transformations from the frame $K$ to the
frame $K'$ and then from the frame $K'$ to the frame
${\widetilde K}_0$ lead to the result
\begin{equation}
   \widetilde {A}^* = L({\bf v}_0') L({\bf v})\, A.
\end{equation}
It is easy to show (see below) that
four-vectors $A^*$ and $\widetilde {A}^*$
do not coincide when the velocities
${\bf v}_0'$ and ${\bf v}$ are not colinear.
Their time components are equal to each
other, however the three-vector ${\bf A}^*$ is turned with
respect to the
three-vector $\widetilde {{\bf A}}^*$ by some angle $\omega$
around the vector $[{\bf v}_0' {\bf v}]$ \cite{bgmr68}. Thus,
\begin{equation}
|{\bf A}^*| = |\widetilde {{\bf A}}^*|,~~
|A_0^*| = |\widetilde {A}_0^*|,~~
          {\bf A}^* = \hat{D}_z(\omega)\widetilde {{\bf A}}^*,
\end{equation}
where $\hat{D}_z(\omega)$ is the matrix of an active
rotation by the angle
$\omega$ around the axis $z$ parallel to the vector
$[{\bf v}_0' {\bf v}]$ (the rotation in Eq.~(6) recovers
the one introduced in Ref. \cite{bgmr68} after the
substitutions
$\hat{D}_z(\omega)\rightarrow
\hat{D}_{z}^{-1}(\omega)$ and $z\rightarrow -z$).
This means that spatial axes of the frames ${\widetilde K}_0$
and $K_0$ are generally not parallel - the transitivity
of parallelism is violated in the theory of relativity
\cite{shi99}.

To find out the rotation angle $\omega$,
let us follow reference \cite{bgmr68}
and assume that the four-vector $A$
is a four-velocity of the frame $K$, {\it i.e.} in this frame
$A=\{c,0,0,0\} $.
At the direct transition from the frame $K$ to the
frame $K_0$, we have from Eq.~(4):
\begin{equation}
          {\bf A}^* = - \gamma_2 {\bf v}_0
\end{equation}
while, at the successive Lorentz transformations
$K \rightarrow K' \rightarrow {\widetilde K}_0$,
we get from Eq.~(5):
\begin{equation}
     \widetilde {{\bf A}}^* = - \gamma_0 \widetilde{{\bf v}}_0.
\end{equation}
Recall that the velocities ${\bf v}_0$ and $\widetilde{{\bf v}}_0$
represent the results of relativistic addition of the velocities
${\bf v}_0'$ and ${\bf v}$ in direct and reverse order given
in Eqs.~(2) and (3), respectively.
According to Eq.~(6),
\begin{equation}
-{\bf v}_0 = - \hat{D}_z(\omega) \widetilde{{\bf v}}_0.
\end{equation}
Consequently,  the angle of the spatial rotation $\omega$ is equal
to the angle between the velocities ${\bf v}_0$ and
$\widetilde{{\bf v}}_0$. Therefore,
$$
  \sin \omega = \frac{|[{\bf v}_0 \widetilde{{\bf v}}_0]|} {v_0^2},
  \quad \cos \omega = \frac{{\bf v}_0 \widetilde{{\bf v}}_0}
  {v_0^2},
$$
where $v_0 = |{\bf v}_0| = |\widetilde{{\bf v}}_0|$.
The positive sign of the angle $\omega $ ($ 0 \le \omega \le \pi$)
is fixed by the negative projection of the difference vector
$\widetilde{{\bf v}}_0-{\bf v}_0$ on the direction of the vector
${\bf v}$ at nonrelativistic velocities.
Using Eqs.~(2) and (3), one can
show \cite{bgmr68} that
\begin{equation}
\sin \omega = \gamma \gamma_0' \frac{v v_0'}{c^2} \sin \theta
\frac{ 1 + \gamma + \gamma_0' + \gamma_0 }
{(1 + \gamma) (1 + \gamma_0')
(1 + \gamma_0)},
\end{equation}
where
$$
\gamma_0 = \widetilde{\gamma}_0=
\gamma \gamma_0' \left (1 + \frac{v v_0'\cos \theta} {c^2}\right ),
$$
$v = |{\bf v}|$, $v_0' = |{\bf v}_0'|$, $\theta$ is the angle
between the vectors  ${\bf v}_0'$ and ${\bf v}$
($ 0 \le \theta \le \pi$).
Taking into account the equality
\begin{equation}
  ( 1 + \gamma + \gamma_0' + \gamma_0)^2 =
2\,(1 + \gamma) (1 + \gamma_0') (1 + \gamma_0) - (\gamma_0^2 - 1)
(\gamma_0'^2 - 1) \sin^2 \theta,
\end{equation}
one can express the cosine of the spatial
rotation angle in the following forms:
 \begin{equation}
 \cos \omega = \frac{(1 + \gamma + \gamma_0' + \gamma_0)^2}
{(1 + \gamma) (1 + \gamma_0') (1 + \gamma_0)} - 1 =
1 - \frac{(\gamma -1)
(\gamma_0' - 1)} {1 + \gamma_0} \sin^2 \theta.
 \end{equation}
Clearly, in the case of colinear velocity vectors
${\bf v}$ and ${\bf v}_0'$ ($\sin\theta=0$),
the angle of the relativistic spatial rotation is equal to zero.

\section{Relativistic spin rotation}
\noindent

The vector of the spin polarization in some frame is constructed
by the Lorentz transformation of the four-tensor of the angular
momentum $ M_{lm}$, or the spin four-vector
$ S_m = (1/2) e_{mlns} M_{lm} u_s$,
from this frame into the \underline{rest frame} of a particle
($ e_{mnls}$ is the antisymmetric unit four-tensor of the fourth
range, $u$ is the particle four-velocity.
Thus, the spin state of a particle depends
on the concrete frame from which the transformation to the rest
frame of this particle is performed.  As a result,
the relativistic spin rotation is a \underline{particular case} of
the kinematical spatial rotation considered above and, it is
described by the same relations.
It should be noted that this fact was not
indicated in the book \cite{bgmr68}, in which the spatial rotation
at successive Lorentz transformations was considered in detail.
On the other hand, in papers [5-7] the effect of the relativistic
spin rotation was discussed without the connection
with the general kinematical effect of the relativistic spatial
rotation.

Let  us assume that $K_0$ is the particle rest frame and
consider the transformations of the spin
four-vector $S$ satisfying the condition $ Su=0$.
In the frame $K$ the spin four-vector is
$S$, the particle velocity is ${\bf v}_0$, and,
in accordance with the Lorentz transformation from
the frame $K$ to the rest frame $K_0$, the spin polarization vector
$\mbox{\boldmath $\zeta$}$
in the frame $K$ (better to say - related to the frame $K$)
is given by the relation
 \begin{equation}
S^* = L({\bf v}_0)S
= \{ 0, \mbox{\boldmath $\zeta$}\}.
\end{equation}

Note that $\mbox{\boldmath $\zeta$}=\langle \hat{\bf s}\rangle$
represents
a mean value of the spin operator $\hat{\bf s}$ in the particle
rest frame. It is related to the normalized spin polarization:
${\bf P}=\mbox{\boldmath $\zeta$}/s$ ($|{\bf P}|\le 1$),
where $s$ is the particle spin.

The spin four-vector $S'$ in the frame $K'$ is connected with the
spin four-vector $S$ in the frame $K$ by the Lorentz
transformation:  $ S' = L({\bf v}) S$, where ${\bf v}$ is
the velocity of the frame $K'$ with respect to the frame $K$.
In accordance with Eq.~(5), we find the spin polarization vector
in the
frame $K'$ as a result of the transition from the frame $K'$
to the rest frame ${\widetilde K}_0$:
\begin{equation}
\widetilde{S}^* = L({\bf v}_0')  L({\bf v})S
= \{ 0, \widetilde{\mbox{\boldmath $\zeta$}}\},
\end{equation}
where ${\bf v}_0'$ is the particle velocity in the frame $K'$.
According to Eq.~(6),
\begin {equation}
   \mbox{\boldmath $\zeta$} = \hat{D}_z(\omega)\,
   \widetilde{\mbox{\boldmath$\zeta$}},
\end{equation}
  $|\mbox{\boldmath $\zeta$}| =
 |\widetilde{\mbox{\boldmath $\zeta$}}| $,\,
$\zeta_z = \widetilde{\zeta}_z $ and
$ z \parallel[{\bf v}_0' {\bf v}]$.
We see that the angle $\omega$ of the relativistic spin rotation
around the axis $z$
can be calculated directly with the help of Eqs.~(10) and (12) for
the kinematical spatial angle. The relation (10) was
just given in the Stapp paper \cite{sta56} on the relativistic
theory of polarization phenomena.

Let us consider the dependence of the angle of the relativistic spin
rotation $\omega=\omega(\theta)$ on the angle $\theta$
between the velocities ${\bf v}_0'$ and ${\bf v}$. Without loss
of generality, both angles $\theta$ and $\omega$ can be
considered in the interval $\langle 0,\pi\rangle$.
It follows from Eqs.~(10) and (12) that
$ \omega( 0) = \omega(\pi) = 0.$
The maximal value
   $$
\omega_{{\rm max}} = \omega(\theta_{{\rm m}})=
\arccos \left [1 - \frac {2(\gamma - 1)
(\gamma_0' - 1)} {(\gamma + 1) (\gamma_0'+1)}\right]
 $$
is achieved at
  $$
    \theta_{{\rm m}} = \arccos \left( - \sqrt{\frac {(\gamma - 1)
 (\gamma_0' - 1)}{(\gamma + 1)(\gamma_0' +1)}}\right ) \geq
\frac{\pi}{2}.
$$
Note that $\omega_{{\rm max}}=\frac{\pi}{2}$, $<\frac{\pi}{2}$
and $>\frac{\pi}{2}$ when $(\gamma_0' -1) (\gamma-1) = 8$,
$<8$ and $>8$, respectively.
\\
At nonrelativistic velocities in the frame $K'$ ($ v_0'/c \ll 1,\,
\gamma_0' \approx 1, \gamma_0 \approx \gamma$) the angle of the spin
rotation is very small ($\omega \ll 1$) :
\begin{equation}
 \omega \approx \frac {\gamma}{\gamma + 1} \frac {v_0' v}{c^2} \sin
\theta.
\end{equation}
It follows  from  Eqs.~(10) and (12) that in the
 ultrarelativistic limit, when $\gamma_0' \rightarrow \infty,\\
\gamma_0/\gamma_0' \rightarrow \gamma[1 + (v/c) \cos \theta]$ :
\begin{equation}
\sin \omega \approx \frac{v}{c}\, \sin \theta \, \frac{1 +
\gamma [1 +(v/c)\cos \theta]}{(1 + \gamma)
[1 + (v/c)\cos \theta]}, \quad
\cos \omega \approx 1 - \frac{\gamma - 1}{\gamma [1 + (v/c)
\cos \theta]} \sin^2 \theta.
\end{equation}
At $v/c \approx 1, \gamma \gg 1$ one has
$$
  \omega_{{\rm max}}\approx \pi - 2\sqrt{2/\gamma},\quad
  \theta_{{\rm m}} \approx \pi  - \sqrt{2/\gamma}.
$$
In this case the sharp dependence on the angle $\theta$ in the
vicinity of $\theta = \pi$ with the width $\sim 1/\sqrt{\gamma}$
takes place \cite{rit61}. Outside this vicinity the angle
$\omega$ is equal approximately  to the angle $\theta$.

The relations (17) are valid exactly for massless particles
(photons, neutrinos). In this situation the spin rotation angle
coincides with the aberration angle (the angle between the vectors
${\bf v}_0'$ and ${\bf v}_0$);  then the spin projection of the
particle onto the direction of its momentum (the helicity,
the degree of the circular polarization of light) is
the relativistic invariant \cite{rit61}.

Indeed, the angle between the particle velocity in the frame
$K'$ and the particle velocity in the frame $K$ is determined,
in general, by the relations:
$$
  \sin \alpha = \frac{|[{\bf v}_0' {\bf v}_0]|}{v_0' v_0} =
 \frac{v}{c} \sin \theta \, \frac{\gamma}{\gamma + 1}
\frac{\gamma_0' + \gamma_0} {\sqrt{\gamma_0^2 - 1}},
$$
\begin{equation}
\cos \alpha =  \frac{{\bf v}_0' {\bf v}_0}{v_0' v_0} =
\frac{\gamma_0' \gamma}{\sqrt{\gamma_0^2 -1}}\,
\left[ \frac{v_0'}{c} + \frac{v}{c} \cos \theta
- \frac{v_0'}{c}\, \frac {\gamma - 1}{\gamma} \sin^2 \theta\right].
\end{equation}
One can show that always the angle of the  relativistic spin
rotation $\omega$ at the transformation $K'\rightarrow K$
is less than the angle of the momentum rotation
$\alpha$  around
the same axis parallel to the vector $[{\bf v}_0' {\bf v}]$.
Thus, $ \omega < \alpha$. However, in the limit
$$
v_0'/c \approx 1,\, \gamma_0' \gg 1,\, \gamma_0 =
\gamma \gamma_0' [1 +(v/c)\cos \theta] \gg 1
$$
the relations (18)  pass into the expressions (17) for
massless particles, so then
$
   \omega = \alpha.
$

\section{ Thomas precession and the effect of the relativistic
 spin rotation}
\noindent

The Thomas precession of the spin polarization vector is the
important
effect of the relativistic kinematics; it takes place even in
the absence of the direct dynamical action on the particle spin
\cite{mol72}.
As a typical example of the Thomas precession,
one can consider the precession of the spin of
a charged particle with zero magnetic moment at its
quasi-classical motion in the external magnetic field \cite{lyu80}.

Due to the orthogonality condition $S(t)u(t)= 0$ for
the spin four-vector $S(t)$ and particle four-velocity $u(t)$
at each instant time $t$,
the following equality holds:
\begin{equation}
\label{19}
     dS(t)\,u(t) + S(t)\,du(t) = 0.
\end{equation}
In the absence of a direct dynamical action on the spin,
the increment $dS(t)$ can be only a linear combination of the
four-vectors $S(t)$ and $u(t)$. Since, in the considered case,
the former contribution
is forbidden due to the conservation of $S(t)^2$
($S(t)dS(t)=0 $), the increment $dS(t)$ is proportional to $u(t)$,
with the coefficient fixed by the general relation (\ref{19}).
The spin four-vectors  at the instant times $t$ and $t +dt$
are then connected by the relation:
 \begin{equation}
dS(t)\equiv S(t + dt) - S(t)= - u(t) \left( S(t)du(t) \right ).
\end{equation}
 The spin polarization vector is determined in the
rest frame of a particle at each instant time;
 in so doing the time $t$ is given in the laboratory frame $K$.
The Lorentz transformations from the
laboratory frame $K$ to the instantaneous rest frames $K_0(t)$ and
$K_0 (t + dt)$ lead to the equalities
$$
 S^*(t) = \{ 0, \mbox{\boldmath $\zeta$}(t)\}_{K_0(t)}=
L({\bf v}(t)) S(t)_K;
$$
\begin{equation}
S^*(t+ dt) = \{ 0, \mbox{\boldmath
$\zeta$}(t+ dt)\}_{K_0(t + dt)}=
L\left( {\bf v}(t)+d{\bf v}(t)\right )S(t + dt)_K.
\end{equation}
Now we will consider the connection between the spin
polarization vectors
$ \mbox{\boldmath $\zeta$}(t)$ and $\mbox{\boldmath $\zeta$}(t+ dt)$.
Let us note that,  according to Eq.~(2) for the relativistic
 addition of velocities, the velocity of the frame $K_0(t_1)$ with
respect to the frame $K_0(t)$,
resulting from the addition of the laboratory velocity ${\bf v}(t_1)$
of the frame $K_0(t_1)$ and the velocity $-{\bf v}(t)$ of the
laboratory frame $K$ with respect to the frame $K_0(t)$, is equal to
\begin{equation}
 \Delta {\bf v}' = \left[\frac{{\bf v}(t_1)}{\gamma(t_1)} -
{\bf v}(t)\left (1 -\frac{{\bf v}(t) {\bf v}(t_1)}{c^2}
\frac {\gamma
(t)}{\gamma(t) +1}\right )\right]
\left ( 1 - \frac{{\bf v}(t) {\bf
v}(t_1)}{c^2}\right)^{-1}.
\end{equation}
At the infinitesimal
time increment one has
$$
t_1 = t +dt,\quad {\bf v}(t_1) = {\bf v}(t) +d{\bf v}.
$$
Then it follows from Eq.~(22) that
\begin{equation}
d{\bf v}' = \gamma(t) d{\bf v} + {\bf v}\, \frac{{\bf v}\,
d{\bf v}}{c^2} \, \frac {\gamma^3(t)}{\gamma(t) + 1}.
\end{equation}
According to the second formula (21), the spin polarization vector
of a particle at
the instant time $t + dt$ is a result of the direct Lorentz
transformation of the spin four-vector at the instant time $t + dt$
from the laboratory frame $K$ to the rest frame
of a particle $K_0(t + dt)$. On the other hand, taking into account
Eq.~(20), one can write
$$
  L({\bf v}(t)) S(t + dt)_K =
 \{ -S(t) \,du, \mbox{\boldmath $\zeta$}(t)\}_{K_0(t)}.
$$
Applying the infinitesimal Lorentz transformation $L(d{\bf v}')$
to both sides of this equation, it is easy to see that,
in the first order in $dt$,
one arrives at the following equality:
$$
  L(d{\bf v}')\, L({\bf v})S(t + dt)_K =
 \{ 0, \mbox{\boldmath $\zeta$}(t)\}_{{\widetilde K}_0(t + dt)}.
$$
Thus, the spin polarization vector at the instant time $t$
can be presented as a result of the successive Lorentz
transformations firstly from the laboratory frame to the frame
$K_0(t)$ and then, from the frame $K_0(t)$ to the rest frame
${\widetilde K}_0(t + dt)$.
In accordance with Eqs.~(6) and (15), we obtain
$$
  L({\bf v}(t) + d{\bf v}) = \hat{D}_z(d\,\omega) L(d{\bf v}')
L({\bf v}(t));
 $$
\begin{equation}
\mbox{\boldmath $\zeta$}(t+ dt) =  \hat{D}_z (d\,\omega)
  \mbox{\boldmath $\zeta$}(t).
\end{equation}
In Eq.~(24) $\hat{D}_z(d\omega)$ is the matrix of the
infinitesimal spatial rotation by the angle $d\,\omega$
around the instantaneous axis $z$ parallel to the vector
$[d {\bf v}'\, {\bf v}]$.

We deduce from Eq.~(24) that the spin polarization vector
rotation for the infinitesimal time interval due the
Thomas precession \underline{coincides with the relativistic spin
rotation} at the transition from the frame $K_0(t)$, in which the
particle velocity is $d{\bf v}'$, to the laboratory frame
$K$, in which the particle velocity is ${\bf v}(t) + d{\bf v}$.

To find the angular velocity of the Thomas precession one should
perform the following substitutions in Eq.~(10) for the angle of
relativistic spatial rotation:
$$
{\bf v}_0' \rightarrow d{\bf v}', \gamma_0' \rightarrow 1, \,
{\bf v} \rightarrow
{\bf v}(t), \, \gamma \rightarrow \gamma (t), \, {\bf v}_0
\rightarrow {\bf v}(t) + d{\bf v},\, \gamma_0 \rightarrow \gamma(t).
$$
Taking into account Eq.~(23) for $d{\bf v}'$, we obtain the
following expression for the infinitesimal angle of the rotation of
the spin polarization vector at the particle motion along the
curvilinear trajectory (in the vector form):
\begin{equation}
d\mbox{\boldmath$\omega$} = \frac{\gamma^2(t)}{\gamma(t) + 1}\,
\frac{1}{c^2}\, [d{\bf v}\, {\bf v}(t)].
\end{equation}
Hence the spin polarization vector satisfies the
precession equation:
\begin{equation}
\frac{d \mbox{\boldmath$\zeta$}(t)}{dt}=
[ \mbox{\boldmath$\Omega$}_{{\rm Th}}(t)\,
\mbox{\boldmath$\zeta$}(t) ]
\end{equation}
with the angular velocity vector of the Thomas precession
\cite{mol72}:
\begin{equation}
\mbox{\boldmath$\Omega$}_{{\rm Th}}(t) \equiv
\frac{d\mbox{\boldmath$\omega$}(t)}{dt} =
-\frac{\gamma^2(t)}{\gamma(t) + 1}\, \frac{1}{c^2}\, [{\bf v}(t)\,
\frac{d{\bf v}(t)}{dt} ].
\end{equation}
It can be
presented also as \cite{lyu80}
\begin{equation} \mbox{\boldmath$\Omega$}_{{\rm Th}}(t) =
- (\gamma(t) - 1) \mbox{\boldmath$\Omega$}_0(t),
\end{equation}
where
$$
\mbox{\boldmath$\Omega$}_0(t) = \left [{\bf l}(t)\,
\frac {d{\bf l}(t)}{dt}\right]
$$
is the instantaneous angular
velocity of the rotation of the particle momentum, ${\bf l}(t)$ is
the unit vector along the momentum direction. According to
Eq.~(28), the direction of the angular velocity of the Thomas
precession is opposite to the direction of the angular velocity of
the momentum rotation.  It is clear that in the case of rectilinear
motion the Thomas precession of spin is absent.

At nonrelativistic velocities, Eq.~(27) reduces to the formula
\begin{equation}
 \mbox{\boldmath$\Omega$}_{{\rm Th}}(t) =
- \frac{1}{2c^2} \left [{\bf v}(t)\, \frac{d{\bf
v}(t)}{dt}\right],
\end{equation}
which describes, in particular, the Thomas precession
of the gyroscope axis.

\section{  Spin precession in the electromagnetic field}
\noindent

Let us consider the precession of the spin polarization vector
of a relativistic particle of charge $q$, mass $m$ and spin $s$,
moving along a quasiclassical trajectory
in the external electromagnetic field.
Its magnetic moment
\begin{equation}
\mu = \frac{q \hbar}{2mc}gs,
\end{equation}
where
$\hbar$ is the reduced Planck constant and
$g$ is the
gyromagnetic ratio determining the anomalous part of the magnetic
moment
$\mu' =\mu(g -2)/g.$ In particular, in the case of an electron
or a negative
muon:  $q = -|e|,\, s =1/2, \, g\approx 2 + e^2/ (\pi
\hbar c)$.

The angular velocity of the spin precession in the electromagnetic
field, which is contained in the equation
$$
\frac{d \mbox{\boldmath$\zeta$}(t)}{dt}=
[ \mbox{\boldmath$\Omega$}(t)\, \mbox{\boldmath$\zeta$}(t) ],
  $$
can be represented as a sum of two terms:
\begin{equation}
\mbox{\boldmath$\Omega$}(t) =
 \mbox{\boldmath$\Omega$}_{{\rm Th}}(t)+
 \mbox{\boldmath$\Omega$}_{{\rm dyn}}(t).
\end{equation}
Here $ \mbox{\boldmath$\Omega$}_{{\rm Th}}(t)$ is the angular
velocity
of the Thomas precession, which is described by Eq.~(28), and
$ \mbox{\boldmath$\Omega$}_{{\rm dyn}}(t)$ is the angular
velocity of
the dynamical precession conditioned by  the presence of the
nonzero
magnetic field ${\bf H}^*(t)$ in the rest frame of a particle.
In accordance with the known result of nonrelativistic quantum
mechanics,
 \begin{equation}
 \mbox{\boldmath$\Omega$}_{{\rm dyn}}(t) =
 -\frac{1}{\gamma(t)} \,\frac{q g}{2mc}\,{\bf H}^*(t);
 \end{equation}
the inverse Lorentz factor in the expression (32) appears due to the
time contraction in the rest frame in comparison with the laboratory
frame
($dt= \gamma(t)dt^*$). According to the Lorentz transformations
of the laboratory magnetic and electric fields
${\bf H}(t)$ and ${\bf E }(t)$ at the point of the particle location,
we have
\begin{equation}
{\bf H}^*(t) = \gamma(t)\left\{ {\bf H}(t) -\frac{\gamma(t) - 1}
{\gamma(t)} {\bf l}(t)({\bf H}(t){\bf l(t)}) +\left [{\bf E}(t)
\frac{{\bf v}(t)}{c}\right]\right \},
\end{equation}
where ${\bf l}(t) = {\bf v}(t)/v(t)$.
The equation of particle motion
 \begin{equation}
   \frac {d{\bf p}(t)}{dt} = q \left({\bf E}(t) +
   \frac{1}{c}[{\bf v}(t)
\,{\bf H}(t)]\right)
 \end{equation}
determines the instantaneous angular velocity
of the rotation of the particle momentum:
\begin{equation}
 \mbox{\boldmath$\Omega$}_0(t) =
-\frac{q}{mc\gamma (t)}\left \{ {\bf H}(t) -{\bf l}(t)
({\bf H}(t) {\bf l}(t)) + \frac{\gamma^2 (t)}{\gamma^2(t) - 1}
\left [{\bf E}(t) \frac {{\bf v}(t)}{c}\right ] \right \}.
\end{equation}
As a result, using Eqs.~(28), (31)-(33) and (35),
we arrive at the formula of
Bargmann, Michel and Telegdi for the total angular velocity of
the spin precession of the relativistic particle at the presence
of the external electromagnetic field \cite{bmt59,blp89}:
$$
 \mbox{\boldmath$\Omega$}(t) =
 -\frac{q}{2mc}\left \{ \left (g - 2 + \frac{2}{\gamma(t)}\right )
{\bf H}(t) - \frac{\gamma(t) - 1}{\gamma(t)}\,(g - 2)\,{\bf l}(t)
({\bf l}(t) {\bf H}(t))\right \}-
$$
\begin{equation}
-\frac{q}{2mc}\left(  g - \frac{2\gamma(t)}{\gamma(t) + 1}
\right )\left [{\bf E}(t) \frac{{\bf v}(t)}{c}\right].
\end{equation}

\subsection{ Magnetic field}
\noindent

In the case of a constant transverse
homogeneous magnetic field, when
${\bf E} = 0, \, {\bf H} {\bf l} = 0$, it follows from
Eqs.~(35) and (36) that
\begin{equation}
 \mbox{\boldmath$\Omega$}_0 = -\frac{q}{mc\gamma}{\bf H};\quad
 \mbox{\boldmath$\Omega$} = -\frac {q}{mc} \left ( \frac{g - 2}{2}
 + \frac{1}{\gamma}\right ).
\end{equation}
 Then the angular velocity of the spin precession
$\mbox{\boldmath$\Omega$}$ and the angular velocity of the momentum
rotation $\mbox{\boldmath$\Omega$}_0$ at the circular motion
in the magnetic field are connected by the proportionality relation
\begin{equation}
\mbox{\boldmath$\Omega$} =
\left ( \frac {g - 2}{2} \gamma + 1\right )
\mbox{\boldmath$\Omega$}_0.
\end{equation}
In so doing the angular velocity of the spin rotation with
respect to the momentum rotation is determined by the anomalous
magnetic moment:
\begin{equation}
   \Delta \mbox{\boldmath$\Omega$} =  \gamma\, \frac {g - 2}{2}\,
\mbox{\boldmath$\Omega$}_0.
\end{equation}
The formula (39) forms the theoretical basis for the measurement of
the muon anomalous magnetic moment \cite{car99}.
Indeed, at the
Dirac value $g = 2$ the projection of the spin polarization vector
onto the momentum direction would not change in time. Let the
initial spin polarization vector of a muon
$\mbox{\boldmath$\zeta$}(t_0)$ be directed along the momentum ${\bf
p}(t_0)$. After $n$ total turns of the muon in the magnetic
field, the angle between the spin polarization vector
$\mbox{\boldmath$\zeta$}(t_0 + nT)$ and the momentum
${\bf p}(t_0 + nT)=
{\bf p}(t_0)$, where $ T = 2\pi/| \mbox{\boldmath$\Omega$}_0|$ is the
period of the circular motion, becomes equal to
\begin{equation}
\Delta \theta = n \pi\, \gamma\, (g - 2).
\end{equation}

\subsection{ Electric field}
\noindent

In the absence of the magnetic field (${\bf H}=0$) it follows from
Eqs.~(35) and (36) that
\begin{equation}
 \mbox{\boldmath$\Omega$}_0(t) = -\frac{q}{mc}
 \frac{\gamma(t)}{\gamma^2(t) - 1}
\left [{\bf E}(t) \frac {{\bf v}(t)}{c}\right ], \quad
 \mbox{\boldmath$\Omega$}(t) =
-\frac{q}{2mc}\left(  g - \frac{2\gamma(t)}{\gamma(t) + 1}
\right )\left [{\bf E}(t) \frac{{\bf v}(t)}{c}\right].
\end{equation}
In so doing,
\begin{equation}
 \mbox{\boldmath$\Omega$}(t) =
 \left (\frac{1}{2}(g - 2) \frac{\gamma^2(t)- 1}{\gamma(t)} +
\frac{\gamma(t) - 1}{\gamma(t)} \right )
\mbox{\boldmath$\Omega$}_0(t).
\end{equation}
In case of a plane trajectory of a charged particle in the electric
field, the vectors $\mbox{\boldmath$\Omega$}(t)$ and $
\mbox{\boldmath$\Omega$}_0(t)$ have a constant direction along the
normal ${\bf n}$ to the plane of the motion.  Then the angle
of the precession of the spin polarization vector is described
by the formula
\begin{equation}
\theta(t) = \int \limits_0^t \left
  [\frac{1}{2} (g - 2) \frac{\gamma^2(t') - 1}{\gamma(t')} +
\frac{\gamma(t') - 1}{\gamma(t')} \right ]
\frac {d\theta_0(t')}{dt'} dt',
\end{equation}
where $\theta_0(t)$ is the angle between the
initial momentum of the particle and its momentum at the instant time
$t$. In case of a negligible change of the particle kinetic energy
at the motion, the connection between the angle of the spin
precession and the angle of the momentum rotation takes a simple form:
\begin{equation}
\theta
= \left [\frac {1}{2} (g - 2) \frac{\gamma^2 - 1}{\gamma} +
\frac{\gamma - 1}{\gamma} \right]\theta_0.
\end{equation}
In the nonrelativistic case
\begin{equation}
\theta = \frac{1}{2}(g - 1)\frac {v^2}{c^2}\theta_0.
\end{equation}
The relation (43) may be applied to the plane channeling of
positively charged particles (protons, positrons) in bent crystals
\cite{lyu80}.
The distortion of the trajectory of a charged particle at its
motion along the bent channel is conditioned by
the average electric field in the plane of the channel which is
perpendicular to the particle momentum. In so doing, the angle of the
particle deviation $\theta_0$ in Eq.~(43) coincides with the angle of
the bend of the crystal and, the axis of the spin rotation
 is perpendicular to the plane of the bend of the crystal.

\subsection{ Thomas precession and the spin-orbit interaction}
\noindent

The central electric field has the structure
\begin{equation}
     {\bf E} = - \frac {1}{r}\frac{dV(r)}{dr}{\bf r},
\end{equation}
where $r = |{\bf r}|,\, V(r)$ is the central potential. According
to Eqs.~(32) (33) and (45) the angular velocity of the dynamical
spin precession in the central electric field can be presented in the
form
\begin{equation}
 \mbox{\boldmath$\Omega$}_{{\rm dyn}} =
\frac{qg}{2m^2c^2\gamma}\,\frac{1}{r}\, \frac {dV(r)}{dr}{\bf L},
\end{equation}
where ${\bf L} = [{\bf r} {\bf p}]$ is the orbital angular momentum.
On the other hand, it follows from Eqs.~(28), (35) and (45) that
the angular velocity of the Thomas precession is given by the
expression
\begin{equation}
 \mbox{\boldmath$\Omega$}_{{\rm Th}} =
 - \frac{q}{m^2 c^2}\, \frac {1}{\gamma + 1}\,\frac{1}{r}\,
\frac{dV(r)}{dr}\,{\bf L}
\end{equation}
As a result, we have:
$$
 \mbox{\boldmath$\Omega$}_{{\rm Th}} =
- \frac{2}{g}\,\frac {\gamma}{\gamma + 1}
 \mbox{\boldmath$\Omega$}_{{\rm dyn}}.
$$
For the electron, the gyromagnetic ratio $ g\approx 2$ and,
 in the nonrelativistic approximation
($\gamma\approx 1$), the following relation holds:
\begin{equation}
\mbox{\boldmath$\Omega$}_{{\rm  Th}} =
- \frac{1}{2}\, \mbox{\boldmath$\Omega$}_{{\rm dyn}}.
\end{equation}
From the point of view of quantum mechanics, the equality of the
spin precession with  the angular velocity
$\mbox{\boldmath$\Omega$}$
is obtained through the commutation of the spin
operator $\hat{{\bf s}}$  with the interaction Hamiltonian
$$
\hat{H}_{{\rm int}} =
\hbar\hat {{\bf s}}\hat{\mbox{\boldmath$\Omega$}}=
\hbar\hat {{\bf s}}\hat{\mbox{\boldmath$\Omega$}}_{{\rm dyn}}+
\hbar\hat {{\bf s}}\hat{\mbox{\boldmath$\Omega$}}_{{\rm Th}}.
$$
Taking into account Eq.~(49), we arrive at the following expression
for the spin-orbit interaction of electrons in atoms \cite{blp89}:
\begin{equation}
 \hat{H}_{{\rm int}} =\frac{1}{2} \hbar \hat{{\bf s}}
\hat {\mbox {\boldmath$\Omega$}}_{{\rm dyn}}=
-\frac{1}{4}\,\frac{|e|\hbar}{m^2c^2}\,\frac{1}{r}\,
\frac{dV(r)}{dr}\,
\hat {\mbox{\boldmath$\sigma$}}\hat {{\bf L}}.
\end{equation}
Here $ \hat {\mbox{\boldmath$\sigma$}} = 2\hat{{\bf s}}$ is the
Pauli
vector operator, $\hat {{\bf L}}$ is the operator of the orbital
angular  momentum.
 Eq.~(50) contains the additional factor 1/2  due to the
contribution of the Thomas precession which reduces twice the
effect of the direct interaction of the magnetic moment of moving
electron with the electric field in the atom.

The spin-orbit interaction
in atoms leads to the fine splitting of atom levels.

\section {Effect of the relativistic spin rotation on
 spin correlations in a two-particle system}
\noindent

The spin correlations in two-particle quantum systems were
analyzed in detail as a tool allowing one to measure the
space--time characteristics of particle production [18,20-23], to
study the two--particle interaction and the production dynamics (see
[19,20-22] and references therein) and to verify the
consequences of the quantum--mechanical coherence with the help of
Bell--type inequalities \cite{lyu00,ll01}.

The spin state of the system of two particles
in an arbitrary frame
is described by the two-particle density matrix, the elements of
which,
$
       \rho^{(1,2)}_{{m_1 m_1}';\,{m_2 m_2}'},
$
are given in the representation of the spin projections of the first
and second particle in the corresponding rest frames onto the
common coordinate axis $z$ (see, {\it e.g.}, \cite{led99,lll03}).
However, one should take into account the relativistic spin rotation
conditioned by the
additional rotation of the spatial axes at the successive
Lorentz transformations along
noncolinear directions [5-7].
As a result,
the {\it concrete} description of a particle spin state
depends on the frame from which the transition to the
particle rest frame is performed.
Particularly, the total spin composition of the two--particle
state with a nonzero vector of relative momentum
 is generally frame--dependent due to different
relativistic rotation angles of the two spins at the transition
to the frame moving in the direction which is not colinear with
the velocity vectors of both particles.

Usually, it is convenient to consider the spin correlations in the
center-of-mass system (c.m.s.) of the particle pair.
This is natural at the addition of the two--particle total spin
and the relative orbital angular momentum into the conserved total
angular momentum.
In some cases, however, it may be useful to make transition to
the laboratory, {\it e.g.}, in the case when the particle
scatterings are used as their spin analyzers \cite{lp97}.\footnote
{In principle, this transition is not necessary since one can
transform the four--vectors defining the
polarization analyzers first to the pair c.m.s.
and then to the respective particle rest frames.}
Denoting $M_1$, $M_2$ and
${\bf p}_1 ={\bf k}$, ${\bf p}_2 = - {\bf k}$ the masses and
c.m.s. momenta of the two particles, their
respective c.m.s. velocities are
${\bf v}_1 =  c {\bf k}/ \sqrt{{\bf k}^2 +M_1^2c^2}$,
${\bf v}_2 = - c {\bf k}/ \sqrt{{\bf k}^2 +M_2^2 c^2}$.
We denote the corresponding laboratory velocities as
${\bf v}_1'$ and ${\bf v}_2'$,
and - the laboratory velocity of the particle pair as ${\bf v }$.
At the Lorentz transformation
from the c.m.s. of the particle pair to the laboratory frame
with parallel respective spatial axes, the spins
of the first and the second particle (in their respective rest
frames)
rotate in opposite directions around the axis which is parallel
to the vector $[{\bf kv}]$.
The angles of the relativistic spin rotation of the first
($i=1$) and
second ($i=2$) particle are found from Eqs.~(10) and (12)
with the substitutions
$$
{\bf v_0'}\rightarrow {\bf v}_i,~~{\bf v_0}\rightarrow {\bf v}_i'
$$
and similarly for the corresponding Lorentz factors.
Thus,
$$
 \sin \omega_1 =  \gamma \gamma_1 \frac {v v_1}{c^2}
  \sin \theta
\frac { 1 + \gamma + \gamma_1 + \gamma_1'}
{(1 + \gamma) (1 + \gamma_1) (1 + \gamma_1')},
$$
\begin{equation}
\cos\omega_1 = 1 - \frac{(\gamma - 1)(\gamma_1 -1)}
{1 + \gamma_1'}\,\sin^2\theta;
\end{equation}
$$
\sin \omega_2 =- \gamma \gamma_2\frac {v v_2}{c^2}
  \sin \theta
\frac { 1 + \gamma + \gamma_2 + \gamma_2'}
{(1 + \gamma) (1 + \gamma_2) (1 + \gamma_2')};
$$
\begin{equation}
\cos\omega_2 = 1 - \frac{(\gamma - 1)(\gamma_2 -1)}
{1 + \gamma_2'}\,\sin^2\theta.
\end{equation}
Here
\begin{equation}
\gamma_1' = \gamma \gamma_1 \left ( 1 +
\frac{v v_1}{c^2}\cos \theta \right ), \quad
\gamma_2' = \gamma \gamma_2 \left ( 1 -
\frac{v v_2}{c^2}\cos \theta \right );
\end{equation}
$\theta=\theta_1$ is the angle between the vectors ${\bf k}$
(${\bf v}_1$) and ${\bf v}$;
$v_i=|{\bf v}_i|$, $v_i'= |{\bf v}_i'|$,
$v =|{\bf v}|$ and $\gamma_i$, $\gamma_i'$, $\gamma$
are the corresponding Lorentz factors.
In the case of equal--mass particles the relations ${\bf v}_1 =
-{\bf v}_2,\, \gamma_1 = \gamma_2$ hold (but
 $\gamma_1' \ne \gamma_2'$ when $\theta \ne \pi/2 $).

It should be noted that the positive sign of the angle of the spin
rotation corresponds to the direction of the
nearest rotation from the vector ${\bf k}$ to the vector
${\bf v}$.
It is clear that the angle between the vectors $-{\bf k}$
(${\bf v}_2$)   and ${\bf v}$
is equal to $\theta_2 = \theta +\pi$, and $\sin\theta_2 =
-\sin\theta$,
$\cos \theta_2 = -\cos \theta$.
Without loss of generality one can assume that $0 \le \theta
\le \pi$.
Since the spins of the first and second particle rotate in
opposite directions, the corresponding rotation angles
 have the opposite signs:  $\omega_1>0,\, \omega_2 <0$.

In the case of the
colinearity of the vectors ${\bf k}$ and ${\bf v}$, when
$\theta =0$
or $\theta =\pi$, both the rotation angles are equal to zero.
At nonrelativistic velocities $v_i$ in the c.m.s. of the particle
pair the
angles $\omega_i$ of the relativistic spin rotation are very small
and scale with $v_i$ (see Eq.~(16) with the substitutions
$v_0'\rightarrow v_1$ and $v_0'\rightarrow -v_2$).
\\

Taking into account the relativistic spin rotation
at the transition from the two--particle c.m.s.
to the laboratory, the two--particle spin
density matrix is transformed as follows:
\begin{equation}
\hat{\rho}'^{(1,2)} = \hat{D}^{(1)}(\omega_1)
\otimes \hat{D}^{(2)}(\omega_2)
\hat{\rho}^{(1,2)} \hat{D}^{(1)+}(\omega_1)
\otimes \hat{D}^{(2)+}(\omega_2),
\end{equation}
where
\begin{equation}
\hat{D}^{(1)}(\omega_1)= \exp (i\omega_1 {\bf \hat{s}}_1{\bf n}),
\quad
\hat{D}^{(2)}(\omega_2) = \exp (i\omega_2 {\bf \hat{s}}_2{\bf n})
\end{equation}
are the matrices of the active space rotations generated by the
vector spin operators $\hat{{\bf s}}_1$ and
$\hat{{\bf s}}_2$,  ${\bf n}$ is the unit vector
parallel to the direction of the vector $[{\bf kv}]$.

In the case of two spin--1/2 particles, the two-particle
spin density matrix has the structure [18,20-22]:
\begin{equation}
\label{eq7}
\hat{\rho}^{(1,2)} =\frac{1}{4} [\hat{I}^{(1)} \otimes
\hat{I}^{(2)} +
 (\hat{\mbox{\boldmath $\sigma$}}^{(1)} {\bf P}_1)
\otimes \hat{I}^{(2)} + \hat{I}^{(1)} \otimes
(\hat{\mbox{\boldmath $\sigma$}}^{(2)}{\bf P}_2)
+ \sum_{l=1}^3
\sum_{k=1}^3 T_{lk}\hat{\sigma}_l^{(1)} \otimes
\hat{\sigma}_k^{(2)}].
\end{equation}
Here $ \hat{I}$ is the two-row unit matrix,
$\hat{\mbox{\boldmath $\sigma$}}^{(i)}=2\hat{{\bf s}}_i$
are the Pauli vector operators,
$ {\bf P}_i=
 \langle \hat{\mbox{\boldmath $\sigma$}}^{(i)}\rangle$
are the polarization vectors ($i =1, 2 $),
$ T_{lk}= \langle
\hat{\sigma}_l^{(1)} \otimes \hat{\sigma}_k^{(2)}\rangle $
are the components of the correlation tensor, $ \{1,2,3\}
\equiv \{x,y,z\}$.
The left and right indexes of the correlation tensor
correspond to the rest frames of the first ($i=1$) and
second ($i=2$)
particle, respectively.
The corresponding probability to select the particles
with the polarizations $\mbox{\boldmath $\alpha$}^{(i)}$
can be obtained by
the substitution of the matrices $\hat{\sigma}_l^{(i)}$
in the expression (\ref{eq7}) with the corresponding
projections $\alpha_l^{(i)}$.
Particularly, when analyzing the polarization states with the
help of particle decays, the vector analyzing power
$\mbox{\boldmath $\alpha$}^{(i)}=\alpha_i {\bf n}_i$,
where $\alpha_i$ is the decay asymmetry corresponding to
the decay analyzer unit vector ${\bf n}_i$.
As a result [20-22], the correlation between the decay analyzers
is determined by the product of the decay asymmetries and the
trace of the spin correlation tensor
$$T=T_{xx}+T_{yy}+T_{zz}.$$
For example, the angular correlation
${\bf n}_1{\bf n}_2=\cos\theta_{12}$
between the directions of the three--momenta of the decay protons
in the respective
rest frames of two $\Lambda$-hyperons decaying into the channel
$\Lambda \rightarrow p + \pi^-$ with the $P$-odd asymmetry
$\alpha = 0.642$ is described by the normalized probability density
[19-22]
 \begin{equation}
 \label{eq7a}
 W(\cos\theta_{12}) = \frac{1}{2} \left( 1 + \alpha^2\frac{T}{3}
 \cos\theta_{12} \right).
 \end{equation}

Clearly, the structure of both Eq.~(\ref{eq7}) and the corresponding
angular distribution of the spin analyzers (e.g., Eq.~(\ref{eq7a}))
does not depend on the system from which the transitions to the
particle rest frames are performed. The system dependence manifests
only through the relativistic rotations in the successive Lorentz
transformations along noncolinear directions. This circumstance was
not understood in paper \cite{al95},  where the unnecessary
condition of nonrelativistic velocities of
$\Lambda$-particles was required.
\section { Transformations of the spin states of two spin--1/2
particles}
\noindent

The matrices of the space rotations due to the transition from
the c.m.s. of two free spin-1/2 particles
to the laboratory are the following:
\begin{equation}
\label{eq8}
\hat{D}^{(1)}(\omega_1) = \cos \frac{\omega_1}{2} +
 i \hat{\mbox{\boldmath $\sigma$}}^{(1)} {\bf n}
\,\sin \frac{\omega_1}{2},\quad
\hat{D}^{(2)}(\omega_2) = \cos \frac{\omega_2}{2} +
 i \hat{\mbox{\boldmath $\sigma$}}^{(2)} {\bf n}
\,\sin \frac{\omega_2}{2}.
\end{equation}
Selecting the $z$--axis parallel to the direction of the vector
${\bf n}=[{\bf kv}]/|[{\bf kv}]|$,
and the axes $x$ and $y$ in the plane perpendicular to this
vector, the polarization vectors and the spin correlation tensor
transform at the transition to the laboratory
in accordance with the active rotations around the
z--axis by the angles $\omega_1$ and $\omega_2$
for the first and second particle, respectively
(the components of the polarization vectors transform according to
Eq.~(15) with the substitutions:
$\widetilde{\mbox{\boldmath $\zeta$}}\rightarrow {\bf P}_i/2$ and
$\mbox{\boldmath $\zeta$}\rightarrow {\bf P}_i'/2$):
$$P'_{i;x}=P_{i;x}\cos\omega_i - P_{i;y}\sin\omega_i;\quad
  P'_{i;y}=P_{i;y}\cos\omega_i + P_{i;x}\sin\omega_i;\quad
  P'_{i;z}=P_{i;z}$$
$$T'_{xx} = (T_{xx}\cos\omega_1 - T_{yx}\sin\omega_1)\cos\omega_2 -
            (T_{xy}\cos\omega_1 - T_{yy}\sin\omega_1)\sin\omega_2; $$
$$T'_{yy} = (T_{yy}\cos\omega_1 + T_{xy}\sin\omega_1)\cos\omega_2 +
            (T_{yx}\cos\omega_1 + T_{xx}\sin\omega_1)\sin\omega_2;
 \quad T'_{zz} = T_{zz};$$
$$T'_{xy} = (T_{xy}\cos\omega_1 - T_{yy}\sin\omega_1)\cos\omega_2 +
            (T_{xx}\cos\omega_1 - T_{yx}\sin\omega_1)\sin\omega_2; $$
$$T'_{yx} = (T_{yx}\cos\omega_1 + T_{xx}\sin\omega_1)\cos\omega_2 -
            (T_{yy}\cos\omega_1 + T_{xy}\sin\omega_1)\sin\omega_2; $$
$$T'_{xz} = T_{xz} \cos\omega_1 - T_{yz} \sin\omega_1;
 \qquad  T'_{zx} = T_{zx} \cos\omega_2 - T_{zy} \sin\omega_2; $$
\begin{equation}
\label{eq9}
T'_{yz} = T_{yz} \cos\omega_1 + T_{xz} \sin\omega_1; \qquad
T'_{zy} = T_{zy} \cos\omega_2 + T_{zx} \sin\omega_2.
\end{equation}
Particularly, the trace of the spin correlation tensor transforms
at the transition to the laboratory as:
\begin{equation}
\label{eq10}
T' = (T_{xx} + T_{yy}) \cos(\omega_1-
\omega_2) + (T_{xy} - T_{yx}) \sin(\omega_1 - \omega_2) + T_{zz}
\end{equation}
or,
in the case of a symmetric tensor, as:
\begin{equation}
\label{eq11}
T'= T - 2\,(T_{xx} + T_{yy})\,\sin^2 \frac{\omega_1 -\omega_2}{2}.
\end{equation}
So, the c.m.s. trace $T$ in Eq.~(\ref{eq7a}) is substituted by the
laboratory one $T'$ calculated using Eq.~(\ref{eq10}) or
(\ref{eq11}) together with Eqs.~(10) and (12) for the spin
rotation angles.\\

 It was shown [20-22] (see also \cite{jay78,bar91})
that the trace of the
correlation tensor of a system of two spin-1/2 particles is the
following linear combination of the relative fractions of singlet
(the total spin $S=0$) and triplet ($S=1$) states:
\begin{equation}
\label{eq13}
 T=\langle \hat{\mbox{\boldmath $\sigma$}}^{(1)} \otimes
 \hat{\mbox{\boldmath $\sigma$}}^{(2)}\rangle =
\rho_t -3\rho_s, \qquad \rho_t + \rho_s =1.
\end{equation}
When we have the pure singlet state of the particle pair in
its c.m.s. ($\rho_s=1,\,\rho_t=0,\\ T_{lk} = - \delta_{lk},\, T=-3$),
the transformation to the laboratory gives
\begin{equation}
\label{eq14}
   T' =  -3 + 4 \sin^2 \frac{\omega_1 - \omega_2}{2}.
\end{equation}
It follows from Eqs.~(\ref{eq13}) and (\ref{eq14}) that at the
transition to the laboratory the relative fraction
of the singlet state decreases in favor of a triplet state:
\begin{equation}
\label{eq15}
 \rho'_s = \cos^2 \frac{\omega_1 - \omega_2}{2}, \qquad
 \rho'_t = \sin^2 \frac{\omega_1 - \omega_2}{2}.
\end{equation}
Thus the square of the total spin of two free particles
with a nonzero vector of relative velocity is
not a relativistic invariant (see \cite{cs58}).
Introducing the two--particle singlet state:
\begin{equation}
|\psi\rangle_{00} =
 \frac{1}{\sqrt{2}} \left(|+1/2\rangle^{(1)}_z \, |-1/2\rangle^{(2)}_z
- |-1/2\rangle^{(1)}_z \,
|+1/2\rangle^{(2)}_z \right)
\end{equation}
and the triplet state with the zero projection onto the rotation
axis $z$:
\begin{equation}
|\psi\rangle_{10} =
 \frac{1}{\sqrt{2}} \left(|+1/2\rangle^{(1)}_z \, |-1/2\rangle^{(2)}_z
+ |-1/2\rangle^{(1)}_z \,
|+1/2\rangle^{(2)}_z \right),
\end{equation}
the result in Eq.~(\ref{eq15}) also follows directly
from the matrices of space rotations in Eq.~(\ref{eq8});
the singlet state in the two--particle c.m.s.
is transformed into the following superposition of the singlet
and triplet states in the laboratory:
\begin{equation}
\label{eq17}
|\psi'_s\rangle =
\cos\frac{\omega_1 - \omega_2}{2}|\psi\rangle_{00}
+ i \sin\frac{\omega_1 - \omega_2}{2}|\psi\rangle_{10}.
\end{equation}

Similarly, the transformation of the pure
triplet state $|\psi\rangle_{10}$ in the two--particle
c.m.s.
 ($\rho_s=0,\, \rho_t=1,\,  T_{zz}=  -1, T_{xx} =T_{yy}=1, T=1$)
to the laboratory gives
  \begin{equation}
  T'= 1 - 4 \sin^2\frac{\omega_1 - \omega_2}{2},
  \end{equation}
the corresponding fractions being
  \begin{equation}
  \rho'_s = \sin^2 \frac{\omega_1 - \omega_2}{2}, \qquad
 \rho'_t = \cos^2 \frac{\omega_1 - \omega_2}{2},
  \end{equation}
in accordance with the transformation:
\begin{equation}
\mid\psi'_t\rangle = \cos\frac{\omega_1 -
\omega_2}{2}|\psi\rangle_{10} + i \sin\frac{\omega_1 -
\omega_2}{2}|\psi\rangle_{00}.
\end{equation}

In the case of the unpolarized triplet in the two--particle c.m.s.
($\rho_s=0,\,\\\rho_t = 1,\, T_{lk}= (1/3)\delta_{lk},
\, T=1$ \cite{lp97,ll01}) we have
 \begin{equation}
T' = 1 - \frac{4}{3} \sin^2 \frac{\omega_1 - \omega_2}{2},
\end{equation}
\begin{equation}
 \rho'_s = \frac{1}{3} \sin^2\frac{\omega_1 - \omega_2}{2},\qquad
 \rho'_t = 1 - \frac{1}{3} \sin^2\frac{\omega_1 - \omega_2}{2}.
\end{equation}

Using Eqs.~(10) and (12), it is easy to show that in the case of
two spin-1/2 particles with the same masses ($\gamma_2 = \gamma_1,\,
v_2 = v_1$) the measure of the spin mixing,
$$ \kappa= \sin^2 \frac {\omega_1 - \omega_2}{2},
$$
can be written in the form:
\begin{equation} \kappa = \frac{(v_1 v)^2}{c^2} \sin^2\theta
\left [\left
(\frac {1}{\gamma} + \frac{1}{\gamma_1}\right)^2 + \frac {( v_1
 v)^2}{c^2} \sin^2 \theta \right ]^{-1}.
\end{equation}
The maximum
of the mixing factor $\kappa$ corresponds to the angle
$\theta = \pi/2$.
In the ultrarelativistic limit, when $\gamma_1\gg 1,\,
\gamma\gg 1$
and $ \sin\theta \gg \max (1/\gamma,\, 1/\gamma_1)$, the factor
$\kappa$
approaches unity.  Then the singlet state in the two--particle
c.m.s.
becomes in the laboratory the triplet state
with the zero projection of the
total spin onto the spin rotation axis $z$
and, {\it vice versa}.
\section{Summary}
\noindent

1. The effect of the relativistic spin rotation at the transition
from some frame to another frame is analyzed. The essence of this
effect lies in the fact that in the framework of the formalism of
inhomogeneous Lorentz group the particle spin state is set in the
particle rest frame and, its concrete description depends on the
frame from which the Lorentz transformation to the rest frame is
performed.\\

2. The connection between the relativistic spin rotation and the
relativistic spatial rotation at the successive Lorentz
transformations along noncolinear directions is considered. It is
shown that the relativistic spin rotation is a purely kinematical
effect; the rotation angle coincides with the angle between
the resulting velocities
at the relativistic addition of two velocities in the direct and
reverse order.\\

3. The Thomas precession of the spin polarization vector at the
particle motion along a curvilinear trajectory is studied. It
is shown that Thomas precession is the effect of the
relativistic kinematics connected with the
effect of the relativistic spin rotation; the expression for the
angular velocity of the Thomas precession is derived using the
formula for the angle of the relativistic spin rotation. \\

4. The spin precession of a relativistic charged particle in the
external electromagnetic field is considered. The cases of the
precession in the homogeneous magnetic field and in the electric
field are discussed. It is established that the angular velocity
of the spin precession can be presented as a simple sum of the
angular velocities of the dynamical precession due to the direct
interaction of the particle magnetic moment with the magnetic
field in the instantaneous rest frame and, the kinematical
Thomas precession connected with the motion of a charge particle
under the action of the Lorentz force.\\

5. The effect of the relativistic spin rotation for a system
of two free particles is investigated setting their spin states
in the corresponding particle rest frames. The transition from the
c.m.s. of two particles to the laboratory frame is considered.
It is shown that the angles
of the relativistic spin rotation of two particles are
generally different, except for the case of colinear
vectors of particle velocities in the laboratory.\\

6. As a result, the effect of the relativistic spin rotation leads to
the dependence of the total two-particle spin composition
(the singlet and triplet fractions in the particular case
of spin-1/2 particles)
on the {\it concrete} frame in which the two-particle system
with a nonzero vector of relative velocity
is analyzed; the total spin is not a Lorentz invariant.
The physical origin of this dependence is the violation
of the parallelism of the spatial axes of the particle rest
frames,
except for the case when the Lorentz transformations to these
frames are done along the directions colinear with the
relative velocity (e.g., from the c.m.s. of the two particles). \\

This work was supported by GA Czech Republic, Grant. No. 292/01/0779,
by Russian Foundation for Basic Research, Grants No. 01-02-16230 and
03-02-16210,  and within the Agreements IN2P3-ASCR No. 00-16 and
IN2P3-Dubna No. 00-46.


\end{document}